\documentstyle[11pt,aaspp]{article}
\def\t0{\theta_{\circ}}

\def\be{\begin{equation}}
\def\en{\end{equation}}

\def\msun{M_{\sun}}

\begin{document}
\title {Protoplanetary Disks in the Nearest Star-Forming Cloud:\\ 
Mid-Infrared Imaging and Optical Spectroscopy of MBM 12 Members}
\author{Ray Jayawardhana\altaffilmark{1,2,3}, Scott J. Wolk\altaffilmark{3,4},
David Barrado y Navascu\'es\altaffilmark{5,6},\\ 
Charles M. Telesco\altaffilmark{2,7}, and 
Thomas J. Hearty\altaffilmark{8}}
\altaffiltext{1}{Department of Astronomy, University of California, Berkeley, 601 Campbell Hall, Berkeley, CA 94720; Electronic mail: rayjay@astro.berkeley.edu}
\altaffiltext{2} {Visiting Astronomer, W.M. Keck Observatory, which is
operated as a scientific partnership among the California Institute of
Technology, the University of California, and the National Aeronautics
and Space Administration.}
\altaffiltext{3} {Visiting Astronomer, Kitt Peak National Observatory,
National Optical Astronomy Observatories, which is operated by the Association
of Universities for Research in Astronomy, Inc. (AURA) under cooperative
agreement with the National Science Foundation.}
\altaffiltext{4}{Harvard-Smithsonian Center for Astrophysics, 60 Garden St., 
Cambridge, MA 02138.}
\altaffiltext{5}{Departamento de F\'{\i}sica Te\'orica, C-XI., Universidad 
Aut\'onoma de Madrid, E-28049 Madrid, Spain and Max-Planck-Institut f\"ur Astronomie, 69117 Heidelberg, Germany}
\altaffiltext{6}{Visiting Astronomer, United Kingdom Infrared Telescope, which
is operated by the Joint Astronomy Centre on behalf of the U.K. Particle
Physics and Astronomy Research Council.}
\altaffiltext{7}{Department of Astronomy, University of Florida, Gainesville, 
FL 32611.}
\altaffiltext{8}{Jet Propulsion Laboratory, Pasadena, CA 91109}

\begin{abstract}
The recent identification of several groups of young stars within 100 parsecs 
of the Sun has generated widespread interest. Given their proximity and 
possible age differences, these systems are ideally suited for detailed 
studies of star and planet formation. Here we report on the first investigation
of protoplanetary disks in one such group, the high-latitude cloud MBM 12
at a distance of $\sim$65 pc. We present mid-infrared observations
of the eight candidate pre-main-sequence (PMS) members and the two 
main-sequence (MS) stars in the same line-of-sight which may or may not be 
associated with the group. We have also derived H$\alpha$ and Li line 
widths from medium-resolution optical spectra. We report the discovery of 
significant mid-infrared excess from six PMS stars --LkH$\alpha$ 262, 
LkH$\alpha$ 263, LkH$\alpha$ 264, E02553+2018, RXJ0258.3+1947 and S18 
--presumably due to optically thick circumstellar disks. Our flux 
measurements for the other two PMS stars and the two MS stars are consistent 
with photospheric emission, allowing us to rule out dusty inner disks. The 
disks we have found in MBM 12 represent the nearest known sample of 
very young protoplanetary systems, and thus are prime targets 
for high-resolution imaging at infrared and millimeter wavelengths.
\end{abstract}


\section{Introduction}
Several groups of young stars within 100 parsecs of the Sun have recently 
received much attention as suitable laboratories for detailed studies 
of star and planet formation. Many of these groups, such as the TW Hydrae 
Association at $\sim$55 pc (Kastner et al. 1997) and the $\eta$ Chamaeleontis 
cluster at $\sim$97 pc (Mamajek, Lawson \& Feigelson 1999), are far from
any obvious parent molecular clouds. Thus, their nature, origin, 
and age are still matters of debate. The most likely explanation for their 
isolation is that these stars are somewhat older than the T Tauri stars found 
in well-known star-forming regions (e.g., Taurus-Auriga) and that their parent 
clouds have dispersed rather quickly. 

However, one nearby young star group {\it does} appear to be associated 
with the high-latitude cloud MBM 12 (Magnani, Blitz, \& Mundy 1985). At a 
distance of $\sim$65 pc, MBM 12 (=L1457) is the nearest known molecular cloud 
(Hobbs, Blitz, \& Magnani 1986; Hearty et al. 2000b). 
Based on {\it ROSAT} 
detections and ground-based follow-up optical spectroscopy, Hearty et al. 
(2000a) have identified 8 late-type young stars in MBM12 and two other 
main-sequence stars in the same line-of-sight which may or may not be related.
Recent surveys suggest that additinal low-mass stars may be present in this 
region (Wolk et al. 2000). 

This intermediate-mass molecular cloud may be representative of how groups 
like TW Hydrae and $\eta$ Cha appeared at an earlier epoch, i.e., at 
$\sim$1 Myr. Since MBM 12, which contains only $\sim$30-100 $\msun$ of 
gas, does not appear to be gravitationally bound, it may be breaking up on a 
timescale comparable to the sound-crossing time. Thus, in a few million years, 
the young stars in MBM12 may appear to be isolated objects, not associated 
with any molecular material --similar to the other nearby ``dispersed'' groups.

Whether MBM12 disperses in a few million years or not, it constitutes an 
excellent sample of extremely nearby young stars for investigating the 
earliest stages of protoplanetary disk systems. In particular, comparison of 
disk properties and statistics between MBM12, TW Hydrae Association
(Jayawardhana et al. 1998, 1999a, 1999b) and $\eta$ Cha cluster could provide 
interesting constraints on the diversity and evolution of circumstellar disks 
(Jayawardhana 2000b). Furthermore, if newborn giant planets or brown dwarfs 
exist around MBM 12 stars, it may be possible to image those very-low-mass 
companions directly using adaptive optics on large ground-based telescopes 
(Jayawardhana 2000a).

Here we present the results of the first investigation of protoplanetary 
disks in the MBM 12 group, using mid-infrared imaging and optical spectroscopy.

\section{Data}
\subsection{K-band photometry}
For 9 of the 10 sources in our sample, we extracted K-band (2.17$\mu$m) 
photometry from the 2-Micron All-Sky Survey (2MASS) point-source catalog. 
In all cases, the 2MASS counterparts were found within 2'' of the nominal 
position and have photometric uncertainties less than 0.1 mag. Since the 2MASS 
catalog does not contain photometry for HD 17332, we have estimated its 
$K$ from $V$=6.87, reported in SIMBAD, and standard $V-K$ color (Kenyon \&
Hartmann 1995) for G0V spectral type.

\subsection{L-band imaging}
We obtained L-band images of all 10 stars in the sample at the United
Kingdom InfraRed Telescope (UKIRT) on Mauna Kea, Hawaii, on February 15
and 16, 2000 (UT). We used IRCAM/TUFTI, a 1-5$\mu$m camera with a 
256$\times$256 InSb detector. The plate scale is 0.081''/pix, resulting in
a 20.8'' field of view. The standard stars HD 40335 ($m_L$=6.43) and HD 106965
($m_L$=7.30) were used for flux calibration.

\subsection{N-band imaging}
We observed the targets in the N-band ($\lambda_0 = 10.8 \mu$m) using 
the OSCIR mid-infrared instrument on the 10-meter Keck II 
telescope on November 19 and 21, 1999 (UT). 
OSCIR is a mid-infrared imager/spectrometer built at 
the University of Florida\footnote[1]{Additional information on  OSCIR is 
available on the Internet at www.astro.ufl.edu/iag/.}, using a 128$\times$128 
Si:As Blocked Impurity Band (BIB) detector developed by Boeing. On Keck II, 
OSCIR has a plate scale of 0.062''/pixel, providing a 7.9''$\times$7.9'' 
field of view. Our observations were made using the standard chop/nod 
technique with a chop frequency of 4 Hz and a throw of 8'' in declination. 
Flux calibration was performed using the mid-infrared standards $\alpha$ Ari
($m_N$=-0.80) and $\alpha$ CMi ($m_N$=-0.76).

\subsection{Optical spectroscopy}
We observed all 10 stars using the Ritchey-Chretien Focus Spectrograph
on the Kitt Peak National Observatory 4-meter telescope on September 6
and 7, 2000 (UT). The BL 450 grating at second order with 632 lines/mm
grating and 1.7'' slit yielded vignetted coverage of the 6030-7500 \AA~ 
region at a resolution of 1.4 \AA/2 pixels. The resulting spectra are shown 
in Figure 1. The equivalent line widths for H$\alpha$ and Li 6708 \AA
(unvignetted), listed in Table 1, were derived by fitting a Voigt profile 
to the lines. This method tends to give slightly ($\sim$10\%) 
smaller values for H$\alpha$ than a simple integration of the area under 
the curve, because it is hard to define the line edges; the effect for 
lithium is minimal.

\section{Results}
In Table 1, we present $K$, $L$ and $N$-band magnitudes for the entire sample, 
as well as $K-L$ and $K-N$ colors. (The photometric error was estimated by
monitoring the variation of standard star fluxes throughout the night and by
comparing the results among several standards.)
For all late-type stars, the photospheric
$K-L$ and $K-N$ $\approx$ 0 (Kenyon \& Hartmann 1995). Thus, excess emission 
at mid-infrared wavelengths is an excellent diagnostic of dusty circumstellar
material in close proximity to young stars (e.g., Jayawardhana et al. 1999b). 
In particular, $K-L \gtrsim$ 0.3-0.5 mag (Kenyon \& Hartmann 1990; Edwards, 
Ray \& Mundt 1993) and $K-N \gtrsim$ 1.2 mag (Skrutskie et al. 1990) indicate 
optically thick inner disks and correlate with characteristic T Tauri spectral
line activity. In the case of MBM 12 stars, the $K$ magnitude is not an exact
measure of photospheric emission because it may be slightly affected by 
extinction and also include a contribution from thermal radiation of the inner
disk. However, corrections of both effects will only increase the measured
$K-L$ and $K-N$ color excesses; thus, our disk detection criterion is a 
conservative one. 

The $K-L$ and $K-N$ colors in Table 1 unambiguosly show significant 
mid-infrared excess from six PMS stars --LkH$\alpha$ 262, LkH$\alpha$ 263,
LkH$\alpha$ 264, E02553+2018, RXJ0258.3+1947 and S18. In all six cases,
the colors are consistent with thermal emission from optically thick 
inner disks. The other two PMS stars --RXJ0255.4+2005 and RXJ0306.5+1921--
do not show a measurable mid-infrared excess within the photometric errors,
allowing us to rule out such disks. HD 17332 and RXJ0255.3+1915, the two
main-sequence stars in the line-of-sight to MBM 12, also lack evidence of
warm circumstellar material. 

Figure 1 shows the optical spectra for all 10 stars in the sample. Three of
the objects --LkH$\alpha$ 262, LkH$\alpha$ 263 and E02553+2018-- clearly
show two components in the H$\alpha$ line, and are probably close binaries.
LkH$\alpha$ 264 also appears to have a blended H$\alpha$ line, but 
the two components are not resolved well enough in our 
medium-resolution spectra to measure their line widths separately. Table 2
lists the H$\alpha$ and Li I 6708 \AA~ line widths of single-line objects
while Table 3 gives the line widths of blue and red H$\alpha$ components 
as well as Li I 6708 \AA~ for the double-line sources. The spectral types
given in Table 2 and 3 are based on Hearty et al. (2000a).

An H$\alpha$ equivalent width greater than 10 \AA~ is generally considered to 
be the accretion signature of a classical T Tauri star with a circumstellar 
disk (Herbig \& Bell 1988). A smaller equivalent width signifies chromospheric 
activity, but no accretion. Therefore, mid-infrared excess we measure should 
be correlated with large H$\alpha$ line widths. Indeed, this generally holds 
true for MBM 12 stars with one notable exception: E02553+2018. This object, 
likely a binary as discussed in Section 4, has large mid-infrared excesses 
($K-L$=1.06, $K-N$=2.45) but weak H$\alpha$ emission in both blue (-1.44 \AA) 
and red (-2.64 \AA) components.

\section{Discussion}
We have detected mid-infrared emission from optically thick disks around 
6 of the 8 known PMS candidates associated with MBM 12. Our results confirm, 
and augment, the evidence for protoplanetary material associated with a 
significant fraction of late-type PMS stars. The exact fraction of stars with 
disks appears to vary
from one star-forming region to another. For example, in their Taurus-Auriga 
sample of stars younger than 3 million years (Myr), Skrutskie et al (1990) 
found that roughly 50\% showed mid-infrared excess consistent with optically 
thick disks (also see Wolk \& Walter 1996). On the other hand, using L-band 
observations, Lada et al. (2000) 
estimated that 80\%-85\% of Trapezium stars harbor circumstellar disks, 
confirming earlier suggestions of a high disk fraction for this 
$\sim$1-Myr-old cluster in Orion. The disk fraction we find in MBM 12 --75\%--
falls in the middle of the range reported for other star-forming regions.
Due to uncertainties in the age estimates of young stellar samples, it is not 
yet clear whether the observed differences in disk frequency are due to rapid 
evolution or environment. 

However, it is interesting to compare the disk properties of MBM 12 to those 
of the TW Hydrae Association and the $\eta$ Cha cluster. The two latter groups
appear to be measurably older; age estimates for both groups, using a variety
of techniques, yield $\sim$10 Myr (Jayawardhana et al. 1999b and references
therein; Mamajek, Lawson \& Feigelson 1999). Our previous work has shown
that many of the TW Hya stars have little or no disk emission at 10$\mu$m 
(Jayawardhana et al. 1999b). Even among the five stellar systems with 10$\mu$m 
excesses, most show some evidence of inner disk evolution. The disk around the 
A0V star HR 4796A has an $r \approx 50$ AU central hole in mid-infrared 
images. The SEDs of HD 98800 and Hen 3-600A also suggest possible inner disk 
holes. The modest excess we detected from CD -33$^{\circ}$7795 could well be 
due to a faint companion. Only TW Hya itself appears to harbor an optically 
thick, actively accreting disk of the kind observed in $\sim$1-Myr-old 
classical T Tauri stars; it is the only one with a large H$\alpha$ equivalent 
width (-220 \AA). Among the 12 known members of the $\eta$ Cha cluster, only
2 have H$\alpha$ equivalent widths larger than 10 \AA~ (Mamajek, Lawson \& 
Feigelson 1999), suggestive of disk accretion. Clearly, the frequency of
classical T Tauri disk systems is significantly higher in MBM 12 ($\sim$75\%)
in comparison to TW Hya Association ($\lesssim$20\%) and $\eta$ Cha cluster 
($\lesssim$20\%). These disk fractions, albeit among small samples of stars, 
suggest that few inner disks survive beyond 10 Myr. This timescale may be 
closely related to the planet formation process (Jayawardhana 2000b).

Our medium-resolution spectra have revealed that at least three of the MBM 12 
sources --LkH$\alpha$ 262, LkH$\alpha$ 263, and E02553+2018-- could be close 
binary systems, although wind absorption cannot be ruled out as the cause of
the double-peaked line profiles. If binaries, it may be possible to resolve 
them with adaptive optics instruments on large ground-based telescopes. Given 
the likely small physical separation of these binaries, follow-up astrometric 
observations can yield direct dynamical masses relatively quickly. Therefore,
MBM 12 binary systems may prove to be useful laboratories for testing 
evolutionary models of low-mass stars. It may also be worth searching for 
cirumbinary disks in these systems. The case of E02553+2018 is puzzling, 
because it has a large mid-infrared excess but weak H$\alpha$ line emission. 
We tentatively suggest that E02553+2018 is a candidate for harboring 
circumbinary dust. 

The H$\alpha$ emission line of LkH$\alpha$ 264 also consists of two components.
However, Lago \& Gameiro (1998), who performed time-series analysis of 
high-resolution profiles, suggest that the H$\alpha$ line of that star is 
produced in two distinct regions --an inner, dense region and an outer, more 
extended region. These authors also found the blue component of the line to 
be highly variable, which may account for the smaller line width we measure
($\sim$18 \AA) when compared to previously published values ($\sim$60-100 \AA~;
Hearty et al. 2000a, Lago \& Gameiro 1998). 

The disks we have found around MBM 12 stars (together with TW Hya) may 
represent 
the nearest known sample of optically thick, actively accreting circumstellar 
disks. Thus, they are ideal for detailed studies of the early stages of 
protoplanetary systems. Their proximity offers a spatial resolution twice 
that of other nearby stellar nurseries such as Taurus-Auriga and Chamaeleon. 
In particular, molecular line and continuum observations using (sub)millimeter 
interferometers will allow us to probe the gaseous component and the outer 
disks of these systems. MBM 12 is also a suitable region for sensitive 
searches of brown dwarfs and massive young planets.

\bigskip
We wish to thank the staff of Keck, KPNO and UKIRT for their outstanding 
support. We are grateful to Scott Fisher and Robert Pi\~na for assistance
with Keck observations and to David Ardila, Michiel Hogerheijde and Kevin 
Luhman for useful discussions. RJ holds a Miller Research Fellowship at the 
University of California, Berkeley. This work was supported in part by NASA 
grant NAG5-8299 and NSF grant AST95-20443 to Geoff Marcy. DByN acknowledges 
support from the  Spanish {\it ``Plan Nacional del Espacio''}, under grant 
ESP98--1339-CO2.

\newpage
\begin{table}
\begin{scriptsize}
\begin{center}
\renewcommand{\arraystretch}{1.2}
\begin{tabular}{lccccc}
\multicolumn{6}{c}{\scriptsize TABLE 1}\\
\multicolumn{6}{c}{\scriptsize INFRARED PHOTOMETRY AND COLOR EXCESS\tablenotemark{a}}\\
\hline
\hline
Name & $K$ & $L$ & $N$ & $K-L$ & $K-N$\\
\hline
\multicolumn{6}{c}{\scriptsize Pre-main-sequence stars}\\
\hline
RXJ0255.4+2005 & 8.99 & 8.91 & 8.84 & 0.08 & 0.15\\
LkH$\alpha$262 & 9.66 & 8.84 & 7.02 & 0.82 & 2.64\\
LkH$\alpha$263 & 9.52 & 8.97 & 7.07 & 0.55 & 2.45\\
LkH$\alpha$264 & 8.90 & 8.13 & 5.15 & 0.77 & 3.75\\
E02553+2018    & 8.11 & 7.05 & 5.66 & 1.06 & 2.45\\
RXJ0258.3+1947 & 10.73& 10.43& 8.58 & 0.30 & 2.15\\
S18	       & 9.56 & 8.66 & 7.81 & 0.90 & 1.75\\
RXJ0306.5+1921 & 9.72 & 9.75 & 9.70 & -0.03& 0.02\\
\hline
\multicolumn{6}{c}{\scriptsize Main-sequence stars}\\
\hline
HD17332\tablenotemark{b} & 5.46 & 5.36 & 5.47 & 0.10 & -0.01\\
RXJ0255.3+1915 & 9.02 & 8.99 & 9.06 & 0.03 & -0.04\\
\hline
\end{tabular}
\end{center}
\tablenotetext{a}{\scriptsize The errors in $K, L, N$ photometry are 
$\pm$0.1 mag.}
\tablenotetext{b}{\scriptsize HD 17332 is a known binary. $L$ and $N$ 
reported in the Table is the total for the binary so that a comparison can 
be made with $K$. For HD17332A, we measure $L$=5.92, $N$=5.97
and for HD17332B, we find $L$=6.34, $N$=6.56.}
\end{scriptsize}
\end{table}
\clearpage

\begin{table}
\begin{scriptsize}
\begin{center}
\renewcommand{\arraystretch}{1.2}
\begin{tabular}{lcccc}
\multicolumn{5}{c}{\scriptsize TABLE 2}\\
\multicolumn{5}{c}{\scriptsize EQUIVALENT WIDTHS OF H$\alpha$ AND Li I 6708\AA~ LINES}\\ 
\multicolumn{5}{c}{\scriptsize FOR STARS WITH ONE COMPONENT}\\
\hline
\hline
Name & Spectral Type & W(H$\alpha$)(\AA)\tablenotemark{a} & W(Li)(\AA) & N(obs)\tablenotemark{b}\\
\hline
RXJ0255.4+2005 & K6 & -0.927 $\pm$ 0.08 & 0.31 $\pm$ 0.03 & 5\\
LkH$\alpha$264\tablenotemark{c} & K5& -17.73 $\pm$ 0.19 & 0.41 $\pm$ 0.01 & 2\\
RXJ0258.3+1947 & M5 & -33.79 $\pm$ 0.58 & 0.42 $\pm$ 0.04 & 2\\
S18	       & M3 & -69.22 $\pm$ 1.25 & 0.35 $\pm$ 0.01 & 2\\
RXJ0306.5+1921 & K1 &  0.31 $\pm$ 0.01 & 0.22 $\pm$ 0.02 & 2\\
HD17332A       & G0 & 1.73 $\pm$ 0.02 & 0.16 $\pm$ 0.01 & 2\\
HD17332B       & G5 &  2.45 $\pm$ 0.03 & 0.14 $\pm$ 0.01 & 2\\
RXJ0255.3+1915 & F9 &  2.69 $\pm$ 0.01 & 0.13 $\pm$ 0.01 & 3\\
\hline
\end{tabular}
\end{center}
\tablenotetext{a}{\scriptsize A negative sign denotes emission.}
\tablenotetext{b}{\scriptsize Number of observations.}
\tablenotetext{c}{\scriptsize Possibly two lines, but unresolved.}
\end{scriptsize}
\end{table}

\begin{table}
\begin{scriptsize}
\begin{center}
\renewcommand{\arraystretch}{1.2}
\begin{tabular}{lccccc}
\multicolumn{6}{c}{\scriptsize TABLE 3}\\
\multicolumn{6}{c}{\scriptsize EQUIVALENT WIDTHS OF H$\alpha$ AND Li I 6708\AA~ LINES}\\
\multicolumn{6}{c}{\scriptsize FOR STARS WITH TWO COMPONENTS}\\
\hline
\hline
Name & Spectral Type & W(Blue H$\alpha$)(\AA)\tablenotemark{a} & W(Red H$\alpha$)(\AA)\tablenotemark{a} & W(Li)(\AA) & N(obs)\tablenotemark{b}\\
\hline
LkH$\alpha$262 & M0 & -13.10 $\pm$ 0.25 & -26.64 $\pm$ 1.80 & 0.51 $\pm$ 0.07 & 3\\
LkH$\alpha$263 & M4 & -8.72 $\pm$ 0.36 & -5.02 $\pm$ 0.45  & 0.39 $\pm$ 0.05 & 3\\
E02553+2018    & K4 & -1.44 $\pm$ 0.10 & -2.64 $\pm$ 0.13  & 0.39 $\pm$ 0.01 & 3\\
\hline
\end{tabular}
\end{center}
\tablenotetext{a}{\scriptsize A negative sign denotes emission.}
\tablenotetext{b}{\scriptsize Number of observations.}
\end{scriptsize}
\end{table}
\clearpage
\newpage
\begin{figure}
\plotfiddle{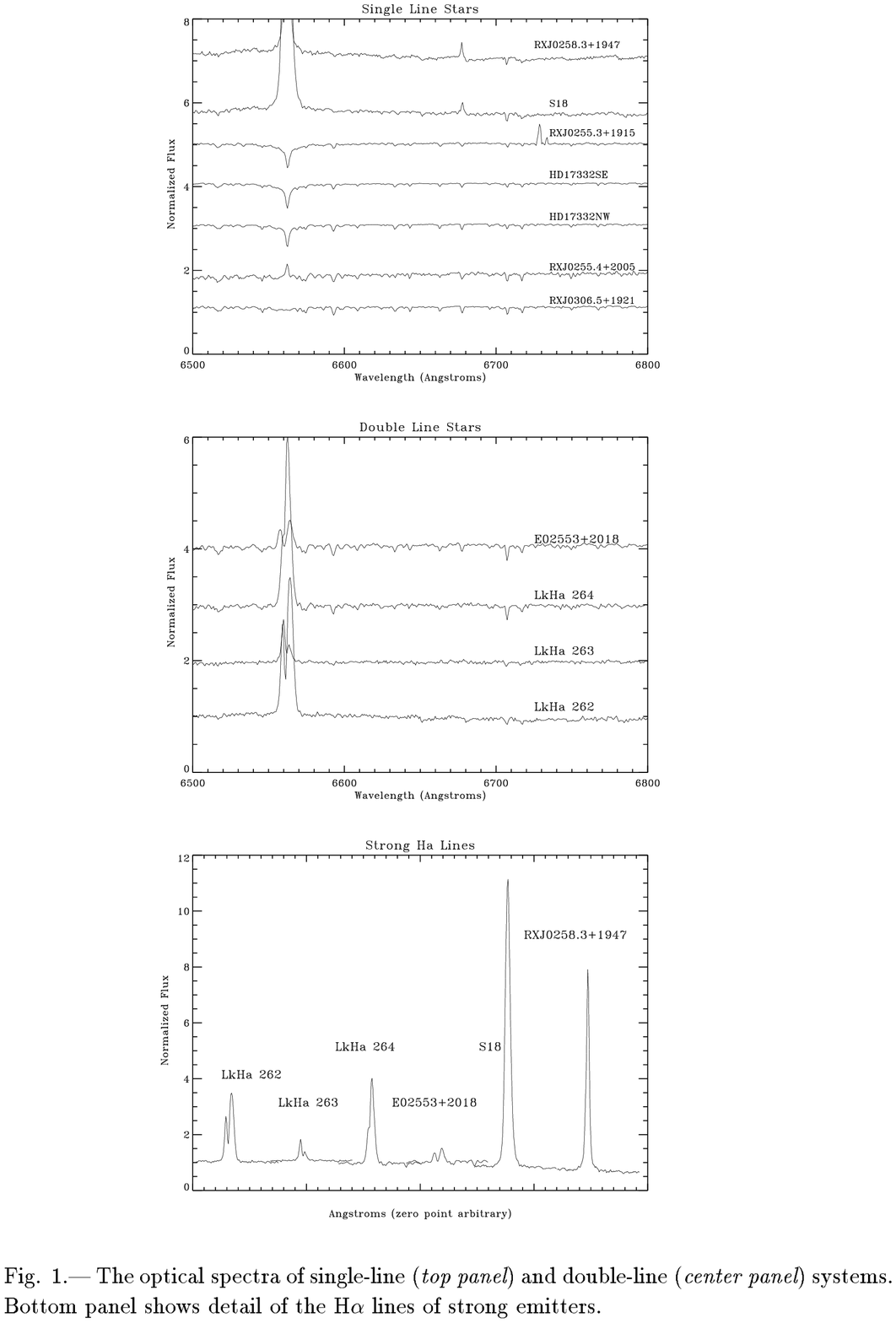}{4in}{0.}{100.}{100.}{-320}{-250}
\end{figure}

\end{document}